\documentclass[aps,prd,nofootinbib]{revtex4}
\pdfoutput=1
\usepackage{epsfig}
\usepackage{amsmath}
\usepackage{graphicx}
\usepackage{multirow}

\begin{document}
\title{ Weak decays of $B_c$ into two hadrons under flavor SU(3) symmetry}

\author{Ruilin Zhu}\email{rlzhu@njnu.edu.cn}
\author{Xin-Ling Han}
\author{Yan Ma }
\author{Zhen-Jun Xiao}\email{xiaozhenjun@njnu.edu.cn}
\affiliation{
 Department of Physics and Institute of Theoretical Physics,
Nanjing Normal University, Nanjing, Jiangsu 210023, China\\
 }

%%%%%%%%%%%%%%%%%%%%
\begin{abstract}
A large number of  $B_c$ meson events have been recorded at the LHCb detector, especially some two-body
hadronic decay modes.  We analyzed the weak decays of the $B_c$ meson into two hadron states  under the
flavor SU(3) symmetry. The relations among amplitudes of $B_c$ into $ D+P(V)$, $B+P(V)$, $P(V)+P(V)$, $T_8+\bar T_8$ and
$T_{10}+\bar T_{10}$
were investigated systematically, where $P$ ( $V$) denotes a light pseudoscalar (vector) meson and $T_{8,10}$ denotes a light baryon.
The $\eta-\eta'$ mixing and $\omega-\phi$ mixing effects are also considered for the phenomenological
discussions. We obtained the relations among decay widths of different $B_c$ decay channels.
These results are helpful to study the  two-body decay properties of the $B_c$ meson and
test the  flavor SU(3) symmetry.
\end{abstract}
\maketitle

%%%%%%%%%%%%%%%%%%%%%%%%%%%%%%%%%%%%%%%%%%%%%%%%%%%%%%%%%%%%%%%%%%%%
\section{Introduction}
%%%%%%%%%%%%%%%%%%%%%%%%%%%%%%%%%%%%%%%%%%%%%%%%%%%%%%%%%%%%%%%%%%%%

The $B_c$ meson family is unique because it is composed of two different heavy flavor quarks, the
charm and bottom. The lifetime  of the $B_c$ meson is greatly longer than that of heavy quarkonia
since it has weak decays only. Of course, the $B_c$ meson's decays become vivid and complicated.
The $B_c$ decays have three kinds of decay modes:
(i) the bottom quark decays through $b\to c, u$, which accounts for around 20 percent to the total decay width;
(ii) the charm quark decays through $\bar{c}\to \bar{s}, \bar{d}$, which accounts for around 70 percent to the
total decay width;
(iii) the weak annihilation, which accounts for around 10 percent to the total decay width~\cite{Kar:2013fna}.

Due to the running of the  Large Hadron Collider (LHC), new experimental data on the decays of $B_c$ meson
are collected, and  several new rare decay channels have been discovered in recent
years~\cite{Aaij:2013cda,Aaij:2014asa,Aaij:2017kea}.
Therein, the $B_c^+\to B_s^0\pi^+$ decay channel by the charm weak transition was first observed by the
LHCb Collaboration~\cite{Aaij:2013cda} and the measured product of the ratio of cross sections
and branching fraction is $[\sigma(B_c^+)/\sigma(B_s^0)]\times Br(B_c^+\to B_s^0\pi^+)
=[2.37\pm0.31(stat)\pm0.11(syst)^{+0.17}_{-0.13}(\tau_{B_c})]\times 10^{-3}$. Except $B_c^+\to B_s^0\pi^+$, other decay channels with the charm weak transition have not been observed currently.
It is worth while to note that a baryonic decay of the $B_c$ meson, $B_c^+\to J/\psi p\bar{p}\pi^+ $,
is observed for the first time, with a significance of $7.3\sigma$~\cite{Aaij:2014asa}.
These measurements will certainly  help us to understand the production and decay properties of  the $B_c$ meson.

Very recently, the LHCb Collaboration have firstly measured the $B_c^+\to D^0 K^+$ decay mode with a
statistical significance of $5.1 \sigma$ using proton-proton collision data corresponding to
an integrated luminosity of 3.0$fb^{-1}$ at 7 and 8 TeV. The ratio between the branching
fraction and that of $B_c^+\to J/\psi \pi^+$ decay mode is given to be ${\cal B}(B_c^+\to D^0 K^+)
/{\cal B}(B_c^+\to J/\psi \pi^+)=0.13\pm0.04(stat)\pm0.01(syst)\pm0.01(R_{J/\psi\pi})$~\cite{Aaij:2017kea}.
For $B_c^+\to D^0 \pi^+$ channel, however, there is no clear event excess in the distribution of the
invariant-mass $m_{D^0\pi^+}$. The LHCb Collaboration only gave the upper limit as
$R_{D^0\pi^+}=(f_c/f_u){\cal B}(B_c^+\to D^0 \pi^+)<3.9\times 10^{-7}$~\cite{Aaij:2017kea}.

Theoretically, the decays of the $B_c$ meson  have been investigated in different approaches.
People employed the theoretical frames such as perturbative QCD (PQCD)
approach~\cite{Du:1988ws,Sun:2008ew,Wen-Fei:2013uea,Liu:2009qa,Liu:2010kq,
Liu:2010nz,Xiao:2013lia,Zou:2017yxc,Rui:2014tpa,Rui:2017pre,Zhang:2009ur,Chang:2017ivy,Chen:2017jrr,Dubnicka:2017job,Rui:2015iia,Rui:2016opu},
QCD sum rules (QCD SR)~\cite{Colangelo:1992cx,Kiselev:1993ea,Azizi:2009ny},
Light-cone sum rules (LCSR)~\cite{Huang:2007kb},
the relativistic quark model (RQM)~\cite{Nobes:2000pm,Ebert:2003cn,Ivanov:2005fd,Ebert:2010zu},
the nonrelativistic constituent quark model (NCQM)~\cite{Hernandez:2006gt},
the light-front quark model (LFQM)~\cite{Wang:2008xt,Wang:2009mi,Ke:2013yka,Shi:2016gqt},
the Bethe-Salpeter equation  method~\cite{Fu:2011tn,Dhir:2008hh},
the nonrelativistic QCD (NRQCD) approach~\cite{Chang:1992pt,Bell:2005gw,Chiladze:1998ny,Qiao:2011yz,Qiao:2012vt,Qiao:2012hp,Zhu:2017lqu,Zhu:2017lwi,Zhu:2018bwp,Wang:2015bka},
Principle of maximum conformality (PMC)\cite{Shen:2014msa}, and Internal and external emission formulae~\cite{Liang:2018rkl}.
From some of references, one can see that different theoretical frames may provide rather different predictions for
the decay width of the same decay channel\cite{Zhu:2017lwi}.
The testing of these theoretical predictions has to refer to future LHCb experiments.

On the other hand, in order to determine the dynamics-independent nature among different decay channels,
the  flavor SU(3)  symmetry approach is a powerful tool to deal with the decays into light hadrons.
Under the  flavor SU(3)  symmetry, the  decay amplitudes are parameterized in terms of SU(3)-irreducible
and model-independent amplitudes. Even though  the size of the amplitudes can not be determined by itself
in the flavor SU(3) symmetry approach, the constraints on certain decay modes are clear.
Thus the flavor SU(3)  symmetry approach has been wildly employed in many works on the weak decays
of heavy flavor mesons and baryons into two or three hadrons
~\cite{Savage:1989ub,Gronau:1995hm,He:1998rq,He:2000ys,Chiang:2004nm,Li:2007bh,
Wang:2009azc,Cheng:2011qh,Hsiao:2015iiu,Lu:2016ogy,He:2016xvd,Wang:2017mqp,Wang:2017azm,
Hu:2017dzi,Shi:2017dto,Wang:2018utj,He:2018php,Wang:2017vnc,Yan:2018gik,Bhattacharya:2017aao}.
The flavor  SU(3)  symmetry approach plays an  important role to bridge dynamic theory
and experimental data in the understanding of the decay properties of the heavy flavor
mesons and baryons.

Up to now, the flavor  SU(3)  symmetry approach has employed to analyze the $B_c$
meson decays into charmed tetraquarks~\cite{He:2016xvd} and charmed or bottomed
mesons~\cite{Bhattacharya:2017aao}.
In this paper, we will systematically explore   $B_c$  decay into $ D+P(V)$, $B+P(V)$, $P(V)+P(V)$,
$T_8+\bar T_8$ and
$T_{10}+\bar T_{10}$
under the flavor $SU(3)$ symmetry respectively, where $P(V)$ denotes the light pseudoscalar(vector) meson
while $T_{8,10}$ denotes a light baryon. In particular we derive relations for  decay widths
and CP violations among different decay channels, which shall be tested by future precise experimental
measurements.

The work is divided into four parts. In Sec.~\ref{sec:II}, we will give an overview  of
flavor $SU(3)$ classification of the hadronic states with different light quarks and their
associated members. In Sec.~\ref{sec:III} we will study the SU(3) decay amplitudes for the
weak $B_c$ decays into two mesons or two baryons.  We will discuss the relations
for decay widths and CP violations in $B_c$ decays in Sec.~\ref{sec:IV}.
We summarize and conclude in the end.

\section{Particle multiplets }\label{sec:II}

Using the standard flavor SU(3) group representation, the $B_c$ meson is a singlet, while the heavy mesons transform as ${\bf 3}$ representation~\cite{Zeppenfeld:1980ex,Chau:1990ay,Gronau:1994rj}, and can be written as $B_i = (B_u(u\bar b), B_d(d \bar b), B_s(s \bar b))$ and $D_i = (D_u(u\bar c), D_d(d \bar c), D_s(s \bar c))$. The light pseudoscalar mesons $P$ with spin-parity $J^P=0^-$ has an octet
\begin{eqnarray}
 P^i_j=\begin{pmatrix}
 \frac{\pi^0}{\sqrt{2}}+\frac{\eta_8}{\sqrt{6}}  &\pi^+ & K^+\\
 \pi^-&-\frac{\pi^0}{\sqrt{2}}+\frac{\eta_8}{\sqrt{6}}&{K^0}\\
 K^-&\bar K^0 &-2\frac{\eta_8}{\sqrt{6}}
 \end{pmatrix}.
\end{eqnarray}
Considering that the SU(3) singlet pseudoscalar state $\eta_1$ can be written as $\left(P_{\eta_1}\right)^i_j=\delta^i_j \eta_1$, thus the physical eigenstates $\eta$ and $\eta'$ can be described by the mixing between  $\eta_1$ and $\eta_8$\footnote{Here we only treated the $\eta$ and $\eta'$  as quark-antiquark configuration, thus the gluonium contribution in the $\eta'$  is not considered. }
\begin{eqnarray}
|\eta\rangle&=&\cos\theta|\eta_8\rangle-\sin \theta|\eta_1\rangle\,,\nonumber\\
|\eta'\rangle&=&\sin\theta|\eta_8\rangle+\cos \theta|\eta_1\rangle\,.
\end{eqnarray}

Similarly, for the light vector meson with spin-parity $J^P=1^-$,  we have the multiplet as
\begin{eqnarray}
 V^i_j=\begin{pmatrix}
 \frac{\rho^0}{\sqrt{2}}+\frac{\omega_8}{\sqrt{6}}  &\rho^+ & K^{*+}\\
 \rho^-&-\frac{\rho^0}{\sqrt{2}}+\frac{\omega_8}{\sqrt{6}}&{K^{*0}}\\
 K^{*-}&\bar K^{*0} &-2\frac{\omega_8}{\sqrt{6}}
 \end{pmatrix}.
\end{eqnarray}
Considering that the SU(3) singlet vector state $\phi_1$ can be written as $\left(V_{\phi_1}\right)^i_j=\delta^i_j \phi_1$, thus the physical eigenstates $\omega$ and $\phi$ can be described by the mixing between  $\phi_1$ and $\omega_8$
\begin{eqnarray}
|\omega\rangle&=&\cos\theta_V|\omega_8\rangle-\sin \theta_V|\phi_1\rangle\,,\nonumber\\
|\phi\rangle&=&\sin\theta_V|\omega_8\rangle+\cos \theta_V|\phi_1\rangle\,,
\end{eqnarray}

The light baryons  with spin-parity $J^P=\frac{1}{2}^+$ form a SU(3) octet $T_8$, which
can be described as
\begin{eqnarray}
\left(T_8\right)^{ij}=\begin{pmatrix}
 \frac{\Sigma^0}{\sqrt{2}}+\frac{\Lambda^0}{\sqrt{6}}  &\Sigma^+ & p^+\\
 \Sigma^-&-\frac{\Sigma^0}{\sqrt{2}}+\frac{\Lambda^0}{\sqrt{6}}&{n}\\
 \Xi^-& \Xi^0 &-2\frac{\Lambda^0}{\sqrt{6}}
 \end{pmatrix}.
\end{eqnarray}

The light baryons  with spin-parity $J^P=\frac{3}{2}^+$ form a SU(3) decuplet $T_{10}$,
the components of  which are
\begin{eqnarray}
&&\left(T_{10}\right)^{111}=\Delta^{++},\quad\quad\quad  \left(T_{10}\right)^{112}=\left(T_{10}\right)^{121}=\left(T_{10}\right)^{211}
=\frac{\Delta^{+}}{\sqrt{3}},\nonumber\\
&&\left(T_{10}\right)^{222}=\Delta^{-},\quad\quad\quad \left(T_{10}\right)^{122}=\left(T_{10}\right)^{211}=\left(T_{10}\right)^{221}
=\frac{\Delta^{0}}{\sqrt{3}},\quad\quad\quad  \left(T_{10}\right)^{333}=\Omega^{-},\nonumber\\
&&\left(T_{10}\right)^{113}=\left(T_{10}\right)^{131}=\left(T_{10}\right)^{311}
=\frac{\Sigma^{'+}}{\sqrt{3}},\quad\quad\quad \left(T_{10}\right)^{223}=\left(T_{10}\right)^{232}=\left(T_{10}\right)^{322}
=\frac{\Sigma^{'-}}{\sqrt{3}},
\nonumber\\
&&\left(T_{10}\right)^{123}=\left(T_{10}\right)^{132}=\left(T_{10}\right)^{213}=\left(T_{10}\right)^{231}
= \left(T_{10}\right)^{312}=\left(T_{10}\right)^{321}
=\frac{\Sigma^{'0}}{\sqrt{6}},\nonumber\\
&&\left(T_{10}\right)^{133}=\left(T_{10}\right)^{313}=\left(T_{10}\right)^{331}
=\frac{\Xi^{'0}}{\sqrt{3}},\quad\quad\quad \left(T_{10}\right)^{233}=\left(T_{10}\right)^{323}=\left(T_{10}\right)^{332}
=\frac{\Xi^{'-}}{\sqrt{3}}.
\end{eqnarray}
From them, one can see the decuplet $T_{10}$ is symmetrical when changing the order of the superscript  $i,j,k$.

%%%%%%%%%%%%%%%%%%%%%%%%%%%%%%%%%%%
\section{SU(3) decay amplitudes for weak $B_c$ decays into two hadrons}\label{sec:III}

\begin{figure}
\begin{center}
\includegraphics[scale=0.6]{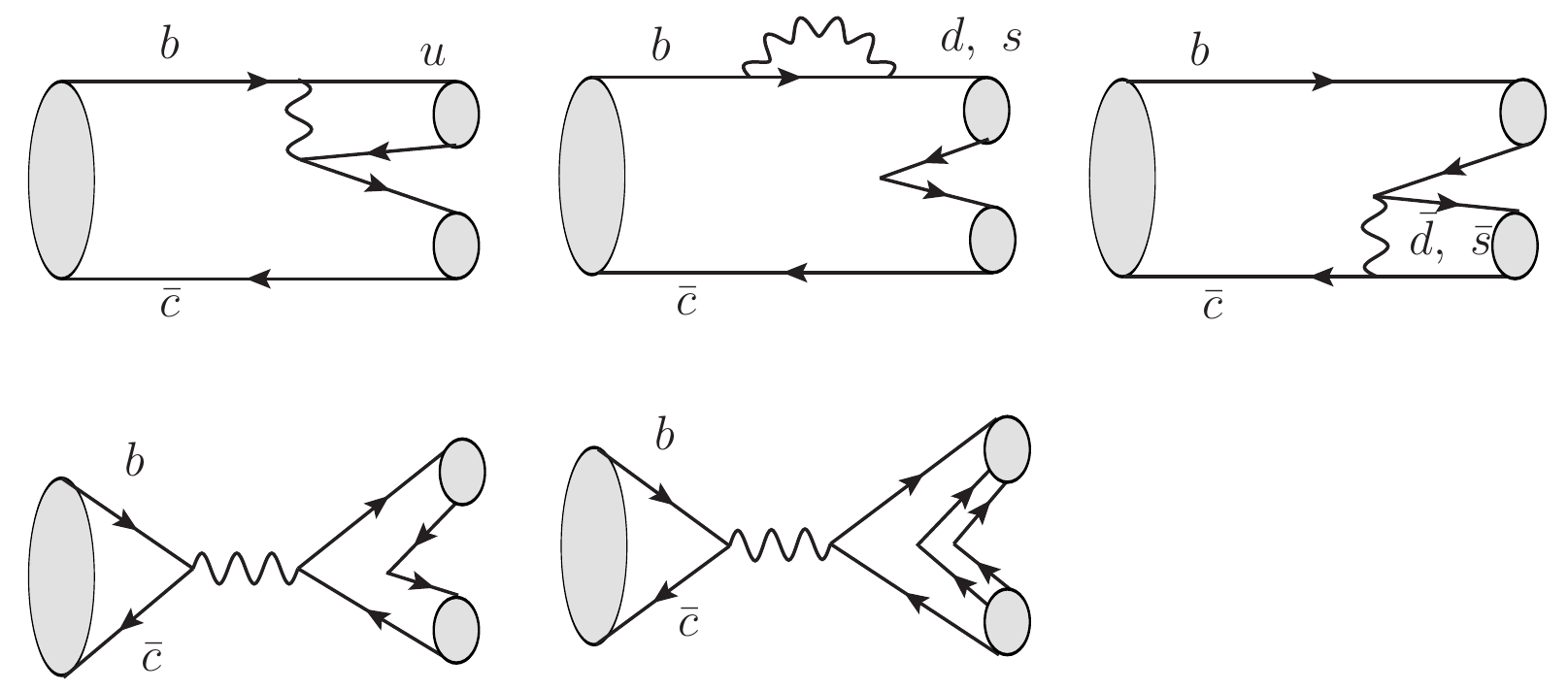}
\end{center}
\caption{Typical Feynman diagrams for the weak $B_c$ decays into two hadrons.  }
\label{fig:feynmand}
\end{figure}

In this section, we study the $B_c\to D_i+P(V)$, $B_c\to B_i+P(V)$, $B_c\to P(V)+P(V)$,
and $B_c\to T_{10}+\bar T_{10}$ decays in the flavor SU(3) symmetry, respectively.
The typical Feynman diagrams for the weak $B_c$ decays into two hadrons  are plotted
in Fig.~\ref{fig:feynmand}. Their decay amplitudes  will be  parameterized in terms of
SU(3)-irreducible amplitudes. They are helpful to get the decay widths relations.

%%%%%%%%%%%%%%%%%%%%%%

\subsection{SU(3) decay amplitudes for $B_c\to D_i+P(V)$ }
First, we will study the bottom quark decays, i.e. $B_c\to D_i+P(V)$  channels. As already mentioned before the $B_c$ is a singlet in the flavor SU(3) group, while the $D_i$ transforms as ${\bf 3}$ representation, and the light pseudo-scalar meson $P$ and vector meson $V$ belong to octets.

The $B_c\to D_q+P(V)$  decays are controlled by the bottom to light quark transition, thus the weak Hamiltonian ${\cal H}_{eff}$ is
 \begin{eqnarray}
 {\cal H}_{eff} &=& \frac{G_{F}}{\sqrt{2}}
     \bigg\{ V_{ub} V_{uq}^{*} \big[
     C_{1}  O^{\bar uu}_{1}
  +  C_{2}  O^{\bar uu}_{2}\big]- V_{tb} V_{tq}^{*} \big[{\sum\limits_{i=3}^{10}} C_{i}  O_{i} \big]\bigg\}+ \mbox{H.c.} ,
 \label{eq:hamiltonian}
\end{eqnarray}
where the $V_{ij}$  is  the CKM matrix element and the $O_{i}$ are the four-fermion effective operators.  According to the group multiplication and decomposition,  we have ${\bf\bar 3}\otimes {\bf3}\otimes {\bf\bar
3}={\bf\bar 3}\oplus {\bf\bar 3}\oplus {\bf6}\oplus {\bf\overline{15}}$. The tree operators is described as  a vector $H^i({\bf\overline3})$, an asymmetrical tensor $H^{[ij]}_k({\bf6})$, and a symmetrical
tensor
$H^{\{ij\}}_k({\bf\overline{15}})$.
From the above formulae, the penguin operators  belong to the ${\bf\bar 3}$ representation.

From the weak Hamiltonian,  the $B_c$ decay amplitudes into a charmed meson $D_i$ and a light pseudo-scalar meson $P$  can be expressed with the  tree  amplitude $A^T_{B_c}$ and the penguin amplitude $A^P_{B_c}$
\begin{eqnarray}
A(B_c\to D_i P) = \langle D_i+P\vert {\cal H}_{eff} \vert B_c \rangle=V_{ub}V^*_{uq}A^T_{B_c} + V_{tb}V^*_{tq}A^P_{B_c}\;,\label{BcA}
\end{eqnarray}
where $q$ denotes the light  down or strange quark. $A^T_{B_c}$ and $A^P_{B_c}$ can be described as the flavor SU(3) amplitudes.
After removing the CKM matrix elements, the  weak Hamiltonian can be rewritten as the sum of the flavor irreducible representations.

For the strangeless decays, i.e. $\Delta S=0(b\to d)$, the flavor irreducible representations have the nonzero
components\cite{Savage:1989ub,He:2000ys,Hsiao:2015iiu}:
\begin{eqnarray}
 H^2({\bf\overline{3}})&=& 1,\;\;\;H^{12}_1({\bf6})=-H^{21}_1({\bf6})=H^{23}_3({\bf6})=-H^{32}_3({\bf6})=1,\nonumber\\
 2H^{12}_1({\bf\overline{15}})&=& 2H^{21}_1({\bf\overline{15}})=-3H^{22}_2({\bf\overline{15}})=
 -6H^{23}_3({\bf\overline{15}})=-6H^{32}_3({\bf\overline{15}})=6.\label{eq:H3615}
\end{eqnarray}
Similarly, one can get the nonzero components of the related  irreducible representations for the $\Delta S=1(b\to s)$
decays. The explicit expressions can be obtained by Eq.~\eqref{eq:H3615}
with the exchange  $2\leftrightarrow 3$.

Thus it is easily to rewrite  the penguin amplitude $A^P_{B_c}$ and the  tree  amplitude $A^T_{B_c}$  as
\begin{eqnarray}
 A^P_{B_c}&=& \alpha H^i({\bf\bar 3}) D_j P^j_i,\nonumber\\
 A^T_{B_c}&=& a_3 H^i({\bf\bar 3}) D_j P^j_i
 + a_6 H^{[ij]}_l ( {\bf6}) D_j P^l_i +  a_{15} H^{\{ij\}}_k({\bf\overline {15}}) D_j P^k_i, \label{eq:amp_Bc_bar6}
\end{eqnarray}
where the $\alpha$ represents the hadronic  parameter from the strong and electro-weak interactions for the penguin topology, while $a_i$ with $i=3,6,15$ represent
the  hadronic parameters from the strong and electro-weak interactions  for the tree topology. Under the flavor SU(3) symmetry, the $B_c$ decay amplitudes into a charmed meson and a light meson can be obtained.
We gave the results in Tab.~\ref{tab:Di+P} and Tab.~\ref{tab:Di+V}. In Tab.~\ref{tab:Di+P},
$\alpha'=\alpha+V_{ub}V^*_{ud}a_3/(V_{tb}V^*_{td})$ and $\alpha^{''}=\alpha+V_{ub}V^*_{us}a_3/(V_{tb}V^*_{ts})$.
In Ref.~\cite{Bhattacharya:2017aao}, the $B_c$ decay amplitudes into a charmed meson
and a light meson are also studied in the flavor SU(3) symmetry. Compared with their
results, we considered the $\eta-\eta'$ mixing effects  and expanding the amplitude relations into the decay channels involving the vector mesons.

%%%%%%%%%%%%%%%%%%%%%%%%%%%%%%%%
\begin{table}
\caption{Decay amplitudes of $B_c\to D_i+P$ decays.  Here and in the following tables, the $\alpha$ and $a_i$ represent the hadronic  parameters.  Besides, $\alpha'=\alpha+V_{ub}V^*_{ud}a_3/(V_{tb}V^*_{td})$ and $\alpha^{''}=\alpha+V_{ub}V^*_{us}a_3/(V_{tb}V^*_{ts})$. }\begin{tabular}{|c|c|c|c|}\hline\hline
channel $\Delta S=0$ & amplitude  \\\hline

$ B_c^-\to D^-\pi^0$ & $\frac{1}{\sqrt{2}}\left(-V_{tb}V^*_{td}\alpha+V_{ub}V^*_{ud}(-a_3+ a_6+5a_{15})\right)$
\\\hline

$ B_c^-\to \bar D^0\pi^-$ &  $V_{tb}V^*_{td}\alpha +V_{ub}V^*_{ud}(a_3-a_6+3a_{15})$
\\\hline

$ B_c^-\to D^-_s K^0$ &  $V_{tb}V^*_{td}\alpha +V_{ub}V^*_{ud}(a_3+ a_6-a_{15})$
\\\hline

$ B_c^-\to D^-\eta$ &  $\frac{1}{\sqrt{6}}\cos \theta\left(V_{tb}V^*_{td}\alpha
+V_{ub}V^*_{ud}(a_3+ 3a_6+3a_{15})\right)-\frac{1}{\sqrt{3}}\sin \theta\,  V_{tb}V^*_{td}\alpha'$
\\\hline

$ B_c^-\to D^-\eta'$ &  $\frac{1}{\sqrt{6}}\sin \theta  \left(V_{tb}V^*_{td}\alpha
+V_{ub}V^*_{ud}(a_3+ 3a_6+3a_{15})\right)+\frac{1}{\sqrt{3}}\cos \theta\,   V_{tb}V^*_{td}\alpha^{''}$
\\

\hline\hline
channel $\Delta S=1$ & amplitude  \\\hline

$ B_c^-\to D^-\bar K^0$ & $V_{tb}V^*_{ts}\alpha +V_{ub}V^*_{us}(a_3+ a_6-a_{15})$
\\\hline

$ B_c^-\to \bar D^0 K^-$ &  $V_{tb}V^*_{ts}\alpha +V_{ub}V^*_{us}(a_3-a_6+3a_{15})$
\\\hline

$ B_c^-\to D^-_s \pi^0$ &  $\sqrt{2} V_{ub}V^*_{us}( a_6+2a_{15})$
\\\hline

$ B_c^-\to D^-_s\eta$ &  $\sqrt{\frac{2}{3}}\cos \theta\left(-V_{tb}V^*_{ts}\alpha
+V_{ub}V^*_{us}(-a_3+3a_{15})\right)-\frac{1}{\sqrt{3}}\sin \theta \,  V_{tb}V^*_{ts}\alpha'$
\\\hline

$ B_c^-\to D^-_s\eta'$ &  $\sqrt{\frac{2}{3}}\sin \theta \left(-V_{tb}V^*_{ts}\alpha
+V_{ub}V^*_{us}(-a_3+3a_{15})\right)+\frac{1}{\sqrt{3}}\cos \theta\,   V_{tb}V^*_{ts}\alpha^{''}$
\\\hline
\end{tabular}\label{tab:Di+P}
\end{table}
%%%%%%%%%%%%%%%%%%%%%%%%%%%%%%

\begin{table}
\caption{Decay amplitudes of $B_c\to D_i+V$ decays. Here and in the following, the $\alpha^V$ and $a^V_i$ represent the hadronic  parameters.  Besides, $\alpha^{V'}=\alpha^{V}+V_{ub}V^*_{ud}a^{V}_3/(V_{tb}V^*_{td})$ and $\alpha^{V''}=\alpha^{V}+V_{ub}V^*_{us}a^{V}_3/(V_{tb}V^*_{ts})$.  }\begin{tabular}{|c|c|c|c|}\hline\hline
channel $\Delta S=0$ & amplitude  \\\hline

$ B_c^-\to D^-\rho^0$ & $\frac{1}{\sqrt{2}}\left(-V_{tb}V^*_{td}\alpha^V +V_{ub}V^*_{ud}(-a^V_3+ a^V_6+5a^V_{15})\right)$
\\\hline

$ B_c^-\to \bar D^0\rho^-$ &  $V_{tb}V^*_{td}\alpha^V +V_{ub}V^*_{ud}(a^V_3-a^V_6+3a^V_{15})$
\\\hline

$ B_c^-\to D^-_s K^{*0}$ &  $V_{cb}V^*_{cd}\alpha^V +V_{ub}V^*_{ud}(a^V_3+ a^V_6-a^V_{15})$
\\\hline

$ B_c^-\to D^-\omega$ &  $\frac{1}{\sqrt{6}}\cos \theta_V \;\left(V_{tb}V^*_{td}\alpha^V
+V_{ub}V^*_{ud}(a^V_3+ 3a^V_6+3a^V_{15})\right)-\frac{1}{\sqrt{3}}\sin \theta_V\; V_{tb}V^*_{td}\alpha^{V'}$
\\\hline

$ B_c^-\to D^-\phi$ &  $\frac{1}{\sqrt{6}}\sin \theta_V \;\left(V_{tb}V^*_{td}\alpha^V
+V_{ub}V^*_{ud}(a_3+ 3a_6+3a_{15})\right)+\frac{1}{\sqrt{3}}\cos \theta_V\; V_{tb}V^*_{td}\alpha^{V''}$
\\

\hline\hline
channel $\Delta S=1$ & amplitude  \\\hline

$ B_c^-\to D^-\bar K^{*0}$ & $V_{tb}V^*_{ts}\alpha^V +V_{ub}V^*_{us}(a^V_3+ a^V_6-a^V_{15})$
\\\hline

$ B_c^-\to \bar D^0 K^{*-}$ &  $V_{tb}V^*_{ts}\alpha^V +V_{ub}V^*_{us}(a^V_3-a^V_6+3a^V_{15})$
\\\hline

$ B_c^-\to D^-_s\rho^0$ &  $\sqrt{2} V_{ub}V^*_{us}( a^V_6+2a^V_{15})$
\\\hline

$ B_c^-\to D^-_s\omega$ &  $\sqrt{\frac{2}{3}}\cos \theta_V\;\left(-V_{tb}V^*_{ts}\alpha^V
+V_{ub}V^*_{us}(-a^V_3+3a^V_{15})\right)-\frac{1}{\sqrt{3}}\sin \theta_V \;V_{tb}V^*_{ts}\alpha^{V'}$
\\\hline

$ B_c^-\to D^-_s\phi$ &  $\sqrt{\frac{2}{3}}\sin \theta_V\;\left(-V_{tb}V^*_{ts}\alpha^V
+V_{ub}V^*_{us}(-a^V_3+3a^V_{15})\right)+\frac{1}{\sqrt{3}}\cos \theta_V\; V_{tb}V^*_{ts}\alpha^{V''}$
\\\hline
\end{tabular} \label{tab:Di+V}
\end{table}

%%%%%%%%%%%%%%%%%%%%%%%%%%%%%%

\subsection{SU(3) decay amplitudes for $B_c\to B_i+P(V)$ }

The second largest decay mode is from the charm decay. According to the estimation of the decay widths, there are three kinds of decay strength:  Cabibbo-allowed by $c\to s\bar{d} u$,
singly Cabibbo-suppressed by $c\to u\bar{d} d/\bar{s} s$, and doubly Cabibbo-suppressed by $c\to d\bar{s} u$.

 We write the nonzero components of the Hamiltonian for the  Cabibbo-allowed decay channels
\begin{eqnarray}
 H^{31}_2({\bf6})=-H^{13}_2({\bf6})=1,\quad\quad\quad
 H^{31}_2({\bf 15})= H^{13}_2({\bf15})=1.
\end{eqnarray}

Combing the $c\to u\bar{d} d$ and $c\to u\bar{s} s$ decays,
 we write the nonzero components of the Hamiltonian for the singly Cabibbo suppressed channels as follows
\begin{eqnarray}
 H^{12}_2({\bf6})=-H^{21}_2({\bf6})=H^{31}_3({\bf6})=-H^{13}_3({\bf6})=\sin \theta_C,\nonumber\\
 H^{31}_3({\bf15})=H^{13}_3({\bf15})=
 -H^{12}_2({\bf15})=-H^{21}_2({\bf15})=\sin \theta_C,\label{eq:H3615}
\end{eqnarray}
where the relation $ V_{ud} V^*_{cd}\simeq  -V_{us} V^*_{cs} \simeq\sin\theta_C $ is employed.

The nonzero components of the Hamiltonian for  the doubly Cabibbo suppressed $c\to d\bar{s} u$ decays are
\begin{eqnarray}
 H^{21}_3({\bf6})=-H^{12}_3({\bf6})=\sin^2 \theta_C,\quad\quad\quad
 H^{21}_3({\bf 15})= H^{12}_3({\bf15})=\sin^2 \theta_C,
\end{eqnarray}
where the relation $ V_{us} V^*_{cd}\simeq |V_{ud} V^*_{cd}|^2 \simeq\sin\theta_C^2 $ is employed.

The decay amplitudes $A(B_c\to B_i+P) = \langle B_i+P\vert {\cal H}_{eff} \vert B_c \rangle$ can be written as $A^T_{B_c}(B_c\to B_i+P)$,  where the representation $H^i({\bf \bar 3})$ will vanish in the
flavor SU(3) symmetry~\cite{He:2018php}.
The effective Hamiltonian can be written as
\begin{eqnarray}
 A^T_{B_c\to B_i+P}&=& a_6 H^{[ij]}_l ( {\bf6}) B_j P^l_i +  a_{15} H^{\{ij\}}_k({\bf\overline {15}}) B_j P^k_i. \label{eq:amp_Bc_bar6}
\end{eqnarray}
The $B_c$ decay amplitudes into a bottom meson and a light meson can be obtained, which are listed in Tab.~\ref{tab:Bi+P} and Tab.~\ref{tab:Bi+V}.

%%%%%%%%%%%%%%%%%%%%%%%%%%%%%%%%
\begin{table}
\caption{Decay amplitudes of $B_c\to B_i+P$ decays.  }\begin{tabular}{|c|c|c|c|}\hline\hline
Cabibbo allowed channel & amplitude  \\\hline

$ B_c^-\to \bar B^0_s \pi^-$ & $V_{cs}V^*_{ud}(-a_6+a_{15})$
\\\hline

$ B_c^-\to B^- K^0$ & $V_{cs}V^*_{ud}(a_6+a_{15})$
\\

\hline\hline
Cabibbo suppressed channel & amplitude  \\\hline
$ B_c^-\to \bar B_s K^-$ & $V_{cs}V^*_{us}(-a_6+a_{15})$
\\\hline

$ B_c^-\to B^- \eta$ & $-\sqrt{\frac{3}{2}}\cos \theta V_{cs}V^*_{us}(a_6+a_{15})$
\\\hline
$ B_c^-\to B^- \eta'$ & $-\sqrt{\frac{3}{2}}\sin \theta V_{cs}V^*_{us}(a_6+a_{15})$
\\\hline
$ B_c^-\to \bar B^0 \pi^-$ & $V_{cd}V^*_{ud}(a_6-a_{15})$
\\\hline
$ B_c^-\to B^- \pi^0$ & $\frac{1}{\sqrt{2}}V_{cd}V^*_{ud}(a_6+a_{15})$
\\
\hline\hline
Doubly Cabibbo suppressed channel & amplitude  \\\hline
$ B_c^-\to \bar B^0 K^-$ & $V_{cd}V^*_{us}(-a_6+a_{15})$
\\\hline

$ B_c^-\to B^- \bar{K}^0$ & $ V_{cd}V^*_{us}(a_6+a_{15})$
\\\hline
\end{tabular} \label{tab:Bi+P}
\end{table}
%%%%%%%%%%%%%%%%%%%%%%%%%%%%%%

%%%%%%%%%%%%%%%%%%%%%%%%%%%%%%%%
\begin{table}
\caption{Decay amplitudes of $B_c\to B_i+V$ decays.  }\begin{tabular}{|c|c|c|c|}\hline\hline
Cabibbo allowed channel & amplitude  \\\hline

$ B_c^-\to \bar B^0_s \rho^-$ & $V_{cs}V^*_{ud}(-a^V_6+a^V_{15})$
\\\hline

$ B_c^-\to B^- K^{*0}$ & $V_{cs}V^*_{ud}(a^V_6+a^V_{15})$
\\

\hline\hline
Cabibbo suppressed channel & amplitude  \\\hline
$ B_c^-\to \bar B_s K^{*-}$ & $V_{cs}V^*_{us}(-a^V_6+a^V_{15})$
\\\hline

$ B_c^-\to B^- \omega$ & $-\sqrt{\frac{3}{2}}\cos \theta_V\; V_{cs}V^*_{us}(a^V_6+a^V_{15})$
\\\hline
$ B_c^-\to B^- \phi$ & $-\sqrt{\frac{3}{2}}\sin \theta_V\; V_{cs}V^*_{us}(a^V_6+a^V_{15})$
\\\hline
$ B_c^-\to \bar B^0 \rho^-$ & $V_{cd}V^*_{ud}(a^V_6-a^V_{15})$
\\\hline
$ B_c^-\to B^- \rho^0$ & $\frac{1}{\sqrt{2}}V_{cd}V^*_{ud}(a^V_6+a^V_{15})$
\\
\hline\hline
Doubly Cabibbo suppressed channel & amplitude  \\\hline
$ B_c^-\to \bar B^0 K^{*-}$ & $V_{cd}V^*_{us}(-a^V_6+a^V_{15})$
\\\hline

$ B_c^-\to B^- \bar{K}^{*0}$ & $ V_{cd}V^*_{us}(a^V_6+a^V_{15})$
\\\hline
\end{tabular}  \label{tab:Bi+V}
\end{table}
%%%%%%%%%%%%%%%%%%%%%%%%%%%%%%
\begin{table}
\caption{Decay amplitudes of $B_c\to P(V)+P(V)$ decays.  }\begin{tabular}{|c|c||c|c|}\hline\hline
channel $\Delta S=0$ & amplitude  & channel $\Delta S=0$ & amplitude  \\\hline

$ B_c^-\to \bar K^0 K^-$ & $V_{cb}V^*_{ud}a_8$ &$ B_c^-\to \bar K^{*0} K^{*-}$ & $V_{cb}V^*_{ud}a^V_8$
\\\hline

$ B_c^-\to \pi^-\eta$ &  $\sqrt{\frac{2}{3}}V_{cb}V^*_{ud}\cos \theta a_8 $ & $ B_c^-\to \rho^-\omega$ &
$\sqrt{\frac{2}{3}}V_{cb}V^*_{ud}\cos \theta_V\; a^V_8 $
\\\hline

$ B_c^-\to \pi^-\eta'$ &  $\sqrt{\frac{2}{3}}V_{cb}V^*_{ud}\sin \theta a_8$ & $ B_c^-\to \rho^-\phi$ &
$\sqrt{\frac{2}{3}}V_{cb}V^*_{ud}\sin \theta_V\; a^V_8 $
\\

\hline\hline
channel $\Delta S=1$ & amplitude   & channel $\Delta S=1$ & amplitude\\\hline

$ B_c^-\to K^- \eta$ & $-\sqrt{\frac{1}{6}}V_{cb}V^*_{us}\cos \theta a_8 $ & $ B_c^-\to K^{*-} \omega$ &
$-\sqrt{\frac{1}{6}}V_{cb}V^*_{us}\cos \theta_V\; a^V_8 $
\\\hline

$ B_c^-\to K^- \eta'$ & $-\sqrt{\frac{1}{6}}V_{cb}V^*_{us}\sin \theta a_8$ &  $ B_c^-\to K^{*-} \phi$ &
$-\sqrt{\frac{1}{6}}V_{cb}V^*_{us}\sin\theta_V\; a^V_8 $
\\\hline

$ B_c^-\to \pi^0 K^-$ &  $\frac{1}{\sqrt{2}}V_{cb}V^*_{us}  a_8$ & $ B_c^-\to \rho^0 K^{*-}$ &
$\frac{1}{\sqrt{2}}V_{cb}V^*_{us}  a^V_8$
\\\hline

$ B_c^-\to \pi^- K^0$ &  $V_{cb}V^*_{us}  a_8$ & $ B_c^-\to \rho^- K^{*0}$ &  $V_{cb}V^*_{us}  a^V_8$
\\\hline
\end{tabular} \label{tab:PV}
\end{table}

\subsection{SU(3) decay amplitudes for $B_c\to P(V)+P(V)$ }

In this subsection and the following subsection, we will study the weak annihilation
accounting for around 10 percent of the $B_c$ total decay width. Let us begin with the
decays into two light mesons, i.e. $B_c\to  P(V)+P(V)$  decay channels. As already mentioned
before  the light pseudo-scalar mesons $P$ and light vector mesons $V$ belong to flavor octets.

The $B_c\to P(V)+P(V)$  decays are induced by the bottom transition into charm quark.
The effective weak Hamiltonian is written as
 \begin{eqnarray}
 {\cal H}_{eff} &=& \frac{G_{F}}{\sqrt{2}}
     \bigg\{ V_{cb} V_{uq}^{*} \big[
     C_{1}  O^{\bar cu}_{1}
  +  C_{2}  O^{\bar cu}_{2}\Big]+ \mbox{H.c.}  \bigg\},
 \label{eq:hamiltonian}
\end{eqnarray}
Since the light quarks in this transition form an octet, we have the nonzero component  for the bottom transition into $c \bar{u} d$
\begin{eqnarray}
 H^2_1({\bf 8})=V_{cb}V_{ud}^*.
\end{eqnarray}
For the bottom transition into $b\to c \bar{u} s$, the non-zero component becomes
\begin{eqnarray}
 H^3_1({\bf 8})=V_{cb}V_{us}^*.
\end{eqnarray}

The decay amplitudes $A(B_c\to P+P) = \langle P+P\vert {\cal H}_{eff} \vert B_c \rangle$ can be expressed as
\begin{eqnarray}
A(B_c\to P+P)&=& a_8 H^{j}_{i}({\bf 8}) P^k_j P^i_k.
\end{eqnarray}
The results of the decay amplitudes of $B_c\to P(V)+P(V)$ can be found  in Tab.~\ref{tab:PV}.

\begin{table}
\caption{Decay amplitudes of $B_c\to T_8+\bar T_8$ decays.  }\begin{tabular}{|c|c||c|c|}\hline\hline
channel $\Delta S=0$ & amplitude  & channel $\Delta S=0$ & amplitude  \\\hline

$ B_c^-\to \Lambda^0 \bar \Sigma^-$ & $\sqrt{\frac{1}{6}} V_{cb}V^*_{ud}a'_8$ &$ B_c^-\to \bar \Lambda^{0} \Sigma^-$  & $\sqrt{\frac{1}{6}}V_{cb}V^*_{ud}a'_8$
\\\hline
$ B_c^-\to \Sigma^0 \bar \Sigma^-$ & $-\sqrt{\frac{1}{2}} V_{cb}V^*_{ud}a'_8$ &$ B_c^-\to \bar \Sigma^{0} \Sigma^-$  & $\sqrt{\frac{1}{2}}V_{cb}V^*_{ud}a'_8$
\\\hline

$ B_c^-\to n\bar p^-$ &  $V_{cb}V^*_{ud} a'_8 $ &-- & --
\\

\hline\hline
channel $\Delta S=1$ & amplitude   & channel $\Delta S=1$ & amplitude\\\hline
$ B_c^-\to \bar \Lambda^0 \Xi^-$ & $\sqrt{\frac{1}{6}} V_{cb}V^*_{us}a'_8$ &$ B_c^-\to \bar \Sigma^{-} \Xi^0$  & $V_{cb}V^*_{us}a'_8$
\\\hline
$ B_c^-\to \bar \Sigma^0\Xi^-$ & $\sqrt{\frac{1}{2}} V_{cb}V^*_{us}a'_8$ &$ B_c^-\to \bar p^{-} \Lambda^0$  & $-\sqrt{\frac{2}{3}}V_{cb}V^*_{us}a'_8$
\\\hline

\end{tabular} \label{tab:T8}
\end{table}

\begin{table}
\caption{Decay amplitudes of $B_c\to T_{10}+\bar T_{10}$ decays.  }\begin{tabular}{|c|c||c|c|}\hline\hline
channel $\Delta S=0$ & amplitude  & channel $\Delta S=0$ & amplitude  \\\hline

$ B_c^-\to \Delta^0 \bar \Delta^-$ & $\frac{2}{3}V_{cb}V^*_{ud}a'_8$ &$ B_c^-\to \Delta^-  \bar \Delta^{0}$  & $\sqrt{\frac{1}{3}}V_{cb}V^*_{ud}a'_8$
\\\hline
$ B_c^-\to \Delta^+ \bar \Delta^{--}$ & $\sqrt{\frac{1}{3}} V_{cb}V^*_{ud}a'_8$ &$ B_c^-\to \Xi^{'-}\bar \Xi^{'0} $  & $\frac{1}{3}V_{cb}V^*_{ud}a'_8$
\\\hline
$ B_c^-\to \Sigma^{'0} \bar \Sigma^{'-}$ & $\frac{\sqrt{2}}{3} V_{cb}V^*_{ud}a'_8$ &$ B_c^-\to \Sigma^{'-} \bar \Sigma^{'0} $  & $\frac{\sqrt{2}}{3}V_{cb}V^*_{ud}a'_8$
\\

\hline\hline
channel $\Delta S=1$ & amplitude   & channel $\Delta S=1$ & amplitude\\\hline
$ B_c^-\to \Sigma^{'-} \bar \Delta^0$ & $\frac{1}{3}V_{cb}V^*_{us}a'_8$ &$ B_c^-\to \Sigma^{'0}  \bar \Delta^{-}$  & $\frac{\sqrt{2}}{3}V_{cb}V^*_{us}a'_8$
\\\hline
$ B_c^-\to \Sigma^{'+} \bar \Delta^{--}$ & $\sqrt{\frac{1}{3}} V_{cb}V^*_{us}a'_8$ &$ B_c^-\to \Xi^{'0}\bar \Sigma^{'-} $  & $\frac{2}{3}V_{cb}V^*_{us}a'_8$
\\\hline
$ B_c^-\to \Omega^{-} \bar \Xi^{'0}$ & $\sqrt{\frac{1}{3}} V_{cb}V^*_{us}a'_8$ &$ B_c^-\to \Xi^{'-} \bar \Sigma^{'0} $  & $\frac{\sqrt{2}}{3}V_{cb}V^*_{us}a'_8$
\\\hline
\end{tabular} \label{tab:T10}
\end{table}

\subsection{SU(3) decay amplitudes for $B_c\to T_8+\bar T_8$ and $B_c\to T_{10}+\bar T_{10}$ }

 In this subsection, we will study the two-body baryonic decays of the $B_c$ meson.
The light baryons with spin-parity $J^P=\frac{1}{2}^+$ form a SU(3) octet $T_{8}$, while the light
baryons with spin-parity $J^P=\frac{3}{2}^+$ form a SU(3) decuplet $T_{10}$.

 The decay amplitudes $A(B_c\to T_8+\bar T_8) = \langle T_8+\bar T_8\vert {\cal H}_{eff} \vert B_c
 \rangle$ can be expressed as
\begin{eqnarray}
A(B_c\to T_8+\bar T_8) =a'_8 H^{j}_{i}({\bf 8}) T^{ki} ({\bf 8})\bar{T}_{jk}({\bf 8})\;.\label{eq:amp_BcT8}
\end{eqnarray}
The corresponding amplitudes results are listed in Tab.~\ref{tab:T8}.

The decay amplitudes $A(B_c\to T_{10}+\bar T_{10}) = \langle T_{10}+\bar T_{10}\vert {\cal H}_{eff} \vert B_c \rangle$
can be expressed as
\begin{eqnarray}
A(B_c\to T_{10}+\bar T_{10}) =a'_8 H^{j}_{i}({\bf 8}) T^{kil}_{10}({\bf 10}) \bar{T}_{jkl}({\bf 10})
\;.\label{eq:amp_BcT10}
\end{eqnarray}
The corresponding amplitudes results are listed in Tab.~\ref{tab:T10}.

\begin{table}
\begin{center}
\caption{ The decay width ratios for $B_c\to B_i+P(V)$. }
\begin{tabular}{|cc|c|c|c|}\hline\hline
&\multicolumn{2}{c|}{$B_c\to B_i+P$}&\multicolumn{2}{c|}{$B_c\to B_i+V$}
\\\hline\hline
&$\Gamma_i/\Gamma(B_c^-\to B^- K^0)$ & $R$  & $\Gamma_i/\Gamma( B_c^-\to B^- K^{*0})$ & $R$
\\\hline
&$\frac{\Gamma(B_c^-\to B^- \pi^0)}{\Gamma(B_c^-\to B^- K^0)}$ & $\frac{|V_{cd}|^2}{2|V_{cs}|^2}$ &
$\frac{\Gamma(B_c^-\to B^- \rho^0)}{\Gamma( B_c^-\to B^- K^{*0})}$ & $\frac{|V_{cd}|^2}{2|V_{cs}|^2}$
\\\hline
&$\frac{\Gamma(B_c^-\to B^- \eta')}{\Gamma(B_c^-\to B^- K^0)}$ &$\frac{ 3|V^*_{us}|^2\sin^2{\theta}}{2|V^*_{ud}|^2}$ &
$\frac{\Gamma(B_c^-\to B^- \phi)}{\Gamma( B_c^-\to B^- K^{*0})}$ &$\frac{ 3|V^*_{us}|^2\sin^2{\theta}}{2|V^*_{ud}|^2}$
\\\hline
&$\frac{\Gamma(B_c^-\to B^- \eta) }{\Gamma(B_c^-\to B^- K^0)}$ &$\frac{ 3|V^*_{us}|^2\cos^2{\theta}}{2|V^*_{ud}|^2}$
& $\frac{\Gamma(B_c^-\to B^- \omega)}{\Gamma( B_c^-\to B^- K^{*0})}$ &$\frac{ 3|V^*_{us}|^2\cos^2{\theta}}{2|V^*_{ud}|^2}$
\\\hline
&$\frac{\Gamma(B_c^-\to B^- \bar{K}^0) }{\Gamma(B_c^-\to B^- K^0)}$ &$\frac{ |V_{cd}V^*_{us}|^2}{|V_{cs}V^*_{ud}|^2}$
& $\frac{\Gamma(B_c^-\to B^- \bar{K}^{*0})}{\Gamma( B_c^-\to B^- K^{*0})}$ &$\frac{ |V_{cd}V^*_{us}|^2}{|V_{cs}V^*_{ud}|^2}$
\\\hline\hline
&$\Gamma_i/\Gamma(B_c^-\to \bar B^0 K^-)$ & $R$
&$\Gamma_i/\Gamma( B_c^-\to \bar B^0 K^{*-})$ & $R$  \\\hline
& $\frac{\Gamma( B_c^-\to \bar B^0 \pi^-)}{ \Gamma(B_c^-\to \bar B^0 K^-)}$ & $\frac{|V^*_{ud}|^2}{|V_{us}^*|^2}$
& $\frac{\Gamma( B_c^-\to \bar B^0 \rho^-)}{\Gamma( B_c^-\to \bar B^0 K^{*-})}$ & $\frac{|V^*_{ud}|^2}{|V_{us}^*|^2}$\\
\hline
&$\frac{ \Gamma(B_c^-\to \bar B_s K^-)}{\Gamma( B_c^-\to \bar B^0 K^-)}$ & $\frac{|V_{cs}|^2}{|V_{cd}|^2}$
&$\frac{\Gamma( B_c^-\to \bar B_s K^{*-})}{\Gamma( B_c^-\to \bar B^0 K^{*-})}$ & $\frac{|V_{cs}|^2}{|V_{cd}|^2}$\\
\hline
\end{tabular} \label{tab:R-Bi}
\end{center}
\end{table}

\begin{table}
\begin{center}
\caption{ The decay width ratios for  $B_c\to P(V)+P(V)$. }
\begin{tabular}{|cc|c|c|c|}\hline\hline
&\multicolumn{2}{c|}{$B_c\to P+P$}&\multicolumn{2}{c|}{$B_c\to V+V$}
\\\hline\hline
&$\Gamma_i/\Gamma(B_c^-\to \bar K^0 K^-)$ & $R$  & $\Gamma_i/\Gamma(B_c^-\to  K^{*0} K^{*-})$ & $R$\\
\hline
& $\frac{\Gamma(B_c^-\to \pi^-\eta)}{\Gamma(B_c^-\to \bar K^0 K^-)}$ & $\frac{2\cos^2{\theta}}{3}$ &
$\frac{ \Gamma(B_c^-\to \rho^-\omega)}{\Gamma(B_c^-\to \bar K^{*0} K^{*-})}$ & $\frac{2\cos^2{\theta}_V}{3}$\\
\hline
&$\frac{\Gamma(B_c^-\to \pi^-\eta')}{ \Gamma(B_c^-\to \bar K^0 K^-)}$ & $\frac{2\sin^2{\theta}}{3}$ &
$\frac{ \Gamma(B_c^-\to \rho^-\phi)}{\Gamma(B_c^-\to \bar K^{*0} K^{*-})}$ & $\frac{2\sin^2{\theta}_V}{3}$\\
\hline\hline
&$\Gamma_i/\Gamma(B_c^-\to \pi^- K^0)$ & $R$
&$\Gamma_i/\Gamma(B_c^-\to \rho^- K^{*0})$ & $R$  \\
\hline
& $\frac{\Gamma(B_c^-\to \pi^0 K^-)}{ \Gamma(B_c^-\to \pi^- K^0)}$ & $\frac{1}{2}$ & $\frac{\Gamma(B_c^-\to
\rho^0 K^{*-})}{ \Gamma(B_c^-\to \rho^- K^{*0})}$ & $\frac{1}{2}$\\
\hline
&$\frac{\Gamma(B_c^-\to K^- \eta')}{\Gamma(B_c^-\to \pi^- K^0)}$ & $\frac{\sin^2{\theta}}{6}$ &
$\frac{ \Gamma(B_c^-\to K^{*-} \phi)}{ \Gamma(B_c^-\to \rho^- K^{*0})}$ &  $\frac{\sin^2{\theta}_V}{6}$\\
\hline
& $\frac{\Gamma(B_c^-\to K^- \eta)}{\Gamma(B_c^-\to \pi^- K^0)}$ & $\frac{\cos^2{\theta}}{6}$ &
$\frac{ \Gamma(B_c^-\to K^{*-} \omega)}{\Gamma(B_c^-\to \rho^- K^{*0})}$ &  $\frac{\cos^2{\theta}_V}{6}$\\
\hline\hline
&$\frac{\Gamma(B_c^-\to \bar K^0 K^-)}{\Gamma(B_c^-\to \pi^- K^0)}$ & $\frac{|V^*_{ud}|^2}{|V^*_{us}|^2}$
& $\frac{\Gamma(B_c^-\to B^- K^{*0})}{\Gamma(B_c^-\to \rho^- K^{*0})}$ & $\frac{|V^*_{ud}|^2}{|V^*_{us}|^2}$\\
\hline
\end{tabular}  \label{tab:R-PV}
\end{center}
\end{table}

\begin{table}
\begin{center}
\caption{  The decay width ratios for $B_c\to T_8+\bar T_8$ and $B_c\to T_{10}+\bar T_{10}$. }
\begin{tabular}{|cc|c|c|c||ccccc}\hline\hline
&\multicolumn{2}{c|}{$B_c\to T_8+\bar T_8$} & \multicolumn{2}{c|}{$B_c\to T_{10}+\bar T_{10}$}
\\\hline
&$\Gamma_i/\Gamma$($ B_c^-\to n\bar p^-$ ) & $R$  & $\Gamma_i/\Gamma$($ B_c^-\to \Xi^{'-}\bar \Xi^{'0} $ ) & $R$\\
\hline
&$\frac{ \Gamma(B_c^-\to \Lambda^0 \bar \Sigma^-) }{\Gamma( B_c^-\to n\bar p^- )}$
& ~~~$\frac{1}{6}$~~~ & $\frac{ \Gamma(B_c^-\to \Delta^0 \bar \Delta^-)}{ \Gamma(B_c^-\to \Xi^{'-}\bar \Xi^{'0}) }$ & $4$
\\\hline
&$\frac{\Gamma( B_c^-\to \bar \Lambda^{0} \Sigma^-)}{\Gamma( B_c^-\to n\bar p^- )}$
& $\frac{1}{6}$ & $\frac{ \Gamma(B_c^-\to \Delta^-  \bar \Delta^{0})}{ \Gamma(B_c^-\to \Xi^{'-}\bar \Xi^{'0} )}$ & $3$\\
\hline
&$\frac{ \Gamma(B_c^-\to \Sigma^0 \bar \Sigma^- )}{\Gamma( B_c^-\to n\bar p^-)}$ &
$ \frac{1}{2}$ & $\frac{ \Gamma(B_c^-\to \Delta^+ \bar \Delta^{--})}{\Gamma( B_c^-\to \Xi^{'-}\bar \Xi^{'0})}$ & $3$\\
\hline
&$\frac{ \Gamma(B_c^-\to \bar \Sigma^{0} \Sigma^- )}{ \Gamma(B_c^-\to n\bar p^- )}$ &
$\frac{1}{2}$ & $\frac{\Gamma( B_c^-\to \Sigma^{'0} \bar \Sigma^{'-})}{ \Gamma(B_c^-\to \Xi^{'-}\bar \Xi^{'0})}$ & $2$\\
\hline
&-& -& $\frac{\Gamma( B_c^-\to \Sigma^{'-} \bar \Sigma^{'0})}{\Gamma( B_c^-\to \Xi^{'-}\bar \Xi^{'0})}$ & $2$
\\\hline\hline
&$\Gamma_i/\Gamma( B_c^-\to \bar \Sigma^{-} \Xi^0)$ & $R$
&$\Gamma_i/\Gamma( B_c^-\to \Sigma^{'-} \bar \Delta^0)$ & $R$  \\
\hline
&$\frac{\Gamma( B_c^-\to \bar \Lambda^0 \Xi^-)}{ \Gamma(B_c^-\to \bar \Sigma^{-} \Xi^0)}$
& $\frac{1}{6}$ &$\frac{ \Gamma(B_c^-\to \Sigma^{'0}  \bar \Delta^{-})}{\Gamma( B_c^-\to \Sigma^{'-} \bar \Delta^0)}$ & $2$\\
\hline
&$\frac{\Gamma(B_c^-\to \bar \Sigma^0\Xi^-)}{\Gamma(B_c^-\to \bar \Sigma^{-} \Xi^0)}$ &
$\frac{1}{2}$ & $\frac{\Gamma(B_c^-\to \Sigma^{'p} \bar \Delta^{--})}{\Gamma( B_c^-\to \Sigma^{'-} \bar \Delta^0)}$ & $3$\\
\hline
&$\frac{\Gamma( B_c^-\to \bar p^{-} \Lambda^0)}{\Gamma(B_c^-\to \bar \Sigma^{-} \Xi^0)}$
& $\frac{2}{3}$ &$\frac{\Gamma(B_c^-\to \Xi^{'0}\bar \Xi^{'-})}{\Gamma(B_c^-\to \Sigma^{'-} \bar \Delta^0)}$ & $4$\\
\hline
& -& -&$\frac{\Gamma(B_c^-\to \Omega^{-} \bar \Xi^{'0})}{ \Gamma(B_c^-\to \Sigma^{'-} \bar \Delta^0})$ & $3$ \\
\hline
&- & -&$\frac{\Gamma(B_c^-\to \Xi^{'-} \bar \Sigma^{'0})}{ \Gamma(B_c^-\to \Sigma^{'-} \bar \Delta^0)}$ & $2$ \\
\hline\hline
& $\frac{\Gamma( B_c^-\to n\bar p^-)}{\Gamma(B_c^-\to \bar \Sigma^{-} \Xi^0)}$ &
$\frac{|V^*_{ud}|^2}{|V^*_{us}|^2}$ &$\frac{\Gamma(B_c^-\to B^- K^{*0})}{\Gamma( B_c^-\to \Sigma^{'-} \bar \Delta^0)}$& $\frac{|V^*_{ud}|^2}{|V^*_{us}|^2}$\\\hline
\end{tabular}  \label{tab:R-T810}
\end{center}
\end{table}

\section{Decay widths relations for weak $B_c$ decays into two hadrons}\label{sec:IV}

Doing a square of  the decay amplitudes and integrating the phase space, their decay widths can be obtained.
Under flavor SU(3) symmetry, there is no difference in the phase space. Thus the relation between decay widths shall be
obtained accordingly. Let us define a ratio between two different channels as
\begin{eqnarray}
R=\Gamma_i/\Gamma_j.
\end{eqnarray}
The results for the weak two-body decays of $B_c$ into hadrons are given in  Table ~\ref{tab:R-Bi}, \ref{tab:R-PV}
and ~\ref{tab:R-T810}. These ratios could  be tested by the future LHCb experiments.

At last, let us study the CP asymmetry effects in the $B_c$ meson decays, which may bring
about the effect at ${\cal O}(10^{-3})$.
From the above amplitudes in tables, the CP conjugated amplitudes can be obtained accordingly.
For example, the CP conjugated amplitudes of the $B_c$ meson decays into a charmed meson and a light meson are
of the form of
\begin{eqnarray}
\frac{\bar{A}(B^+_c\to \bar{D}_i+\bar{P})}{A(B^-_c\to D_i+P)}&=& \frac{V^*_{ub}V_{uq}A^T_{B_c} + V^*_{tb}V_{tq}A^P_{B_c}}{V_{ub}V^*_{uq}A^T_{B_c} + V_{tb}V^*_{tq}A^P_{B_c}}.
\end{eqnarray}
The direct CP asymmetry can then be defined as
\begin{eqnarray}
{\cal A}_{CP}^{dir}&=& \frac{\Gamma(B_c^+\to\bar{f_1}\bar{f_2})-\Gamma(B_c^-\to f_1f_2)}{\Gamma(B_c^+\to\bar{f_1}\bar{f_2})+\Gamma(B_c^-\to f_1f_2)}
=\frac{|\bar{A}|^2-|A|^2}{|\bar{A}|^2+|A|^2}.
\end{eqnarray}
From this equation, the direct CP asymmetry  can be obtained when inputting the corresponding decay amplitudes.
We do not list these repeated results. The  decay channels of the $B_c$ meson into a bottom meson
and a light meson, two light mesons, or two baryons have no direct CP asymmetry,
because only tree topology diagrams contribute.
From the interference of tree diagrams and penguin diagrams, these decay channels
of the $B_c$ meson decays into a charmed meson and a light meson have the direct CP asymmetry.

Employing the unitarity of the CKM matrix with $(VV^\dag)_{ij}=(V^\dag V)_{ij}=\delta_{ij}$, we get
\begin{eqnarray}
Im[V^*_{tb}V_{td}V_{ub}V^*_{ud}]&=& -Im[V^*_{tb}V_{ts}V_{ub}V^*_{us}].
\end{eqnarray}
The relations for the direct CP asymmetry of the $B_c$ meson decays into a charmed meson and a light meson are
\begin{eqnarray}
\frac{{\cal A}_{CP}^{dir}(B_c^-\to \bar D^0\pi^-)}{{\cal A}_{CP}^{dir}(B_c^-\to \bar D^0 K^-)}&=& -\frac{\Gamma(B_c^-\to \bar D^0 K^-)}{\Gamma(B_c^-\to \bar D^0\pi^-)},\\
\frac{{\cal A}_{CP}^{dir}(B_c^-\to D^-_s\bar K^0)}{{\cal A}_{CP}^{dir}(B_c^-\to D^- \bar K^0)}&=& -\frac{\Gamma(B_c^-\to D^- \bar K^0)}{\Gamma(B_c^-\to D^-_s\bar K^0)},\\
\frac{{\cal A}_{CP}^{dir}(B_c^-\to \bar D^0\rho^-)}{{\cal A}_{CP}^{dir}(B_c^-\to \bar D^0 K^{*-})}&=& -\frac{\Gamma(B_c^-\to \bar D^0 K^{*-})}{\Gamma(B_c^-\to \bar D^0\rho^-)},\\
\frac{{\cal A}_{CP}^{dir}(B_c^-\to D^-_s\bar K^{*0})}{{\cal A}_{CP}^{dir}(B_c^-\to D^- \bar K^{*0})}&=& -\frac{\Gamma(B_c^-\to D^- \bar K^{*0})}{\Gamma(B_c^-\to D^-_s\bar K^{*0})}.
\end{eqnarray}
These relations actually are very general and similar to the relations in $B$ meson decays~\cite{Fleischer:1999nz,Gronau:2000zy}, which shall be tested by  future LHCb experiments.

%%%%%%%%%%%%%%%%%%%%%%%%%%%%%%%%
 \section{conclusions}
%%%%%%%%%%%Do we need a conclusion section??? %%%%%%%%%%%%%%%%%%
 \label{sec:conclusion}
%%%%%%%%%%%%%%%%%%%%%%%%%%%%%%%%

In this paper, we investigated the decay width relations for the $B_c$ weak decays
into two hadrons under the flavor SU(3) symmetry.
The corresponding decay amplitudes are described by the summation of the  flavor SU(3)  irreducible
amplitudes. The decay channels of the $B_c$ meson into a charmed meson and a light meson, a bottom meson
and a light meson, two light mesons, or two baryons were studied systematically.
The direct CP asymmetry effects only exist in the decay channels of the $B_c$ meson into a charmed
meson and a light meson.
And we obtained some direct CP asymmetry relations in the  flavor SU(3) symmetry.
Hadron-hadron colliders provide a solid platform and the precision tests for the various decay modes of the double heavy
$B_c$ meson.
The theoretical predictions for  the decay width ratios $R$ and the direct CP asymmetries for the considered
$B_c$ weak decays into two hadrons, as listed in the Tables of this paper,
could be tested in the future LHCb experiments.

\section*{Acknowledgement}

We thank the discussions with Prof. Wei Wang. This work was supported in part by the National Natural Science Foundation
of China under Grant No. 11705092 and 11775117, by Natural Science Foundation of
Jiangsu under Grant No.~BK20171471, and by the start-up funds of Nanjing Normal University.

\end{document}